\documentstyle[12pt]{article}

\begin{document}
\newcommand {\be}{\begin{equation}}
\newcommand {\ee}{\end{equation}}
\newcommand {\bea}{\begin{array}}
\newcommand {\cl}{\centerline}
\newcommand {\eea}{\end{array}}
\newcommand {\f}{\over}
\newcommand {\eps}{\epsilon}

\baselineskip 0.65 cm
\begin{flushright}
IPM -97-247\\
hep-th/9710121
\end{flushright}
\begin{center}
\bf \Large {Classification of Different Branes at Angles}
\end{center}
\cl{\it M.M. Sheikh Jabbari}
\cl {\it Institute for studies in theoretical Physics and Mathematics 
(IPM) } \cl{\it P.O.Box 19395-5746, Tehran,Iran}
\cl{\it Department of physics Sharif University of Technology}
\cl{\it P.O.Box 11365-9161}
\cl{\it e-mail  jabbari@netware2.ipm.ac.ir}
\vskip 2cm
\begin{abstract}

In this paper, we consider two D-branes rotated with respect to each other,
and argue that in  this way one can find brane configurations preserving 
${1 \f 16}$ of SUSY.
Also we classify different brane configurations preserving    
${1 \f 2}$, ${1 \f 4}$, ${3 \f 16}$,${1 \f 8}$, ${1 \f 16}$ of SUSY.

\end{abstract}
\newpage
{\bf 1.  Introduction}
 
 Branes and their different configurations have played crucial role in 
developing dualities of string theories and M-theory. These branes are BPS 
states which preserve half of the available supersymmetry [1]. The possible 
bound state of D-branes with F(undamental)-strings or with themselves or 
$NS_5$-branes [2-4] have been of special interest in testing duality 
conjectures in string theory or low energy effective field theories.

The stability of these bound states is supported by no-force condition induced 
by the supersymmetry preserved by that special configuration of branes. Due to
the no-force conditions these branes form  {\it marginally} bound states.
These marginal bounds have been widely studied through solutions of low energy 
effective SUGRA theories in 10 and 11 dimensions and also by SUSY algebra 
arguments [5-13].

The non-marginal bound states of branes can also be constructed. In string 
theory these are bound states of $D_p$-brane with F-string [3] or $D_5$-branes 
with $NS_5$-branes [2,3] and in M-theory, they are as bounds states of $M_2$ and
$M_5$-branes [14]. All the non-marginal cases are believed to be BPS excited 
states of $D_p$-branes or $NS_5$-branes or $M_5$-branes respectively [15].
These states as discussed in [3,14] preserve ${1 \f 2}$ of the SUSY charges 
present, like individual $D_p$-branes or $NS_5$-branes or $M_5$-branes.

In Ref.[5] p-branes making two or three angles have been studied and shown that
in a special configuration with two angles, i.e. when two angles are equal,
${1 \f 8}$ of SUSY survives.
A more general argument on $M_5$-branes at angles is given by Townsend [16],
where configurations of branes preserving ${3 \f 16}$ and also branes preserving
${1 \f 8}$  of SUSY have been studied through SUSY algebra arguments.

In this paper I find a new family of D-branes at angles which preserve  
${1 \f 16}$ of SUSY. These branes are rotated with respect to each other at 
4-angles. By means of different dualities we can go to M-theory level 
at this level we argue that two $M_5$-branes rotated at 4 anlges can preserve 
${1\f 16}$ of SUSY. Finally we classify all the brane configurations with 
various fractions of SUSY.
This article is organized as following:
              
In section 2, we introduce two $D_4$-branes making 4 angles, in string theory 
. These branes are rotated with respect to each other by rotations of $U(4)$ 
of $SO(8)$ rotation group and by means of open strings stretched between such 
branes, calculate their interactions. Vanishing conditions of this amplitude 
which is a sign of remaining some SUSY, is studied and some special cases which
reproduces the previous works is discussed.

In section 3, I consider other different D-branes ($(p,p')$-branes) at angles and
again study non-interacting configurations.

In section 4, the results of sections 2,3 are checked with SUSY algebra 
arguments. In this way we can generalize our string theoretic results to 
$NS_5$-branes too, and also by means of dualities regenerate results of Ref.  
[16] and find the ${1 \f 16}$ configuration in M-theory.

At last in section 5, We classify these results and gather them in a table which
is organized in fraction of preserved SUSY: ${1 \f 2}$, ${1 \f 4}$, ${3 \f 16}$,
${1 \f 8}$, ${1 \f 16}$.
\vskip 0.5cm 
{\bf 2.   D-branes at angles in string theory}

Consider two $D_p$-branes, world volume of one of them spans $01...p$ space
and the second is rotated with respect to it. For $(p<5)$ in the most general
case they can be described by $p$ angles $(\theta_1,\theta_2,...,\theta_p)$
($0 \leq \theta_i \leq \pi$), which show the Abelian rotations in $U(p)$ 
which is a subspace of ${SO(2p) \f SO(p)\times SO(p)}$ rotation group, 
i.e. $\theta_i$ ($ 1 \leq i \leq p$) shows rotation in $(i,p+i)$ plane.

These branes can be introduced by the following rotated boundary  conditions
on open strings ending on each brane [17]:

\be
\sigma=0 \left\{  \begin{array}{cc}
X^{\mu}=0 & \mu=p+1,..,9 \\ 
\partial_{\sigma}X^{\mu}=0  &  \mu=0,..,p
\end{array}\right.
\ee
\be
\sigma=\pi \left\{  \begin{array}{cc}
X^{\mu}=Y^{\mu} & \mu=2p+1,..,9 \\ 
\partial_{\sigma}X^{\mu}=0  &  \mu=0 \\
X^{i} \sin{\theta_i}+X^{(p+i)}\cos{\theta_i} =0 \\
\partial_{\sigma}X^{i}\cos {\theta_i}-
\partial_{\sigma}X^{(p+i)}\sin {\theta_i}=0  & i=1,...,p.
\end{array}\right.
\ee

So for two $D_2$-branes 2 angles, for $D_3$-branes 3 angles and for 
$D_4$-branes  4 angles are the most general cases. For $(p \geq 5)$ the most general
case is described by $(9-p)$ angles and again all the arguments of p-branes holds  
for $(9-p)$, more precisely the most general case is governed by $min(p, 9-p)$
angles.

It is easy to see that the case which is described by only two (or three) angles
(for arbitrary $p$ greater or smaller than 4) is obtained from 4 angles case by
putting two (or one) of them to zero, so we will focus on the 4 angles case.

Along the calculations given in [17] one can find the mode expansion of the 
$X^{\mu}$ and also the world sheet fermions $\psi^{\mu}_\pm$ for NS and R sectors.
To calculate the corresponding interaction between two branes we should find
amplitude of one closed string exchange, this amplitude can be expressed by
one loop vacuum amplitude of open strings stretched between branes:  
\be 
A=2\int {dt \over 2t}\;\; Tre^{-t H},
\ee 
where H is the open string Hamiltonian. The $Tr$ should be performed on momentum 
modes and oscillator modes, which are allowed with proper GSO projection.
The amplitude then is calculated to be
\be
 A(\theta_i)=2 \int \frac{dt}{2t}(8\pi^2\alpha't)^{-1/2}e^{-\frac{Y^2t}{2\pi^2\alpha'}} 
 (\bf{ NS-R}), 
\ee
where $\bf{ NS,R}$ are given by
\be
{\bf R}=\prod_{j=1}^{4}\frac{\Theta_2(i\theta_j t \mid it)}{\Theta_1(i\theta_j t\mid it)}
-\prod_{j=1}^{4}\frac{\Theta_1(i\theta_j t\mid it)}{\Theta_1(i\theta_j t\mid it)}
\ee
\be
{\bf NS}=\prod_{j=1}^{4}\frac{\Theta_3(i\theta_j t \mid it)}{\Theta_1(i\theta_j t\mid it)}
-\prod_{j=1}^{4}\frac{\Theta_4(i\theta_j t\mid it)}{\Theta_1(i\theta_j t\mid it)}
\ee

As it is proved in Appendix, the following identity between 
${\bf \Theta}$-functions:
\be
\prod_{j=1}^{4}\Theta_3(i\theta_j t \mid it)
-\prod_{j=1}^{4}\Theta_4(i\theta_j t\mid it)
=
\prod_{j=1}^{4}\Theta_2(i\theta_j t \mid it)
-\prod_{j=1}^{4}\Theta_1(i\theta_j t\mid it)
\ee
holds provided:  \footnote {These conditions are upto an integer factor of  
$2\pi$.}  

\be
\theta_1+\theta_2+\theta_3+\theta_4=0 \;\;\ 0r \;\;\ \theta_1\pm \theta_2=
\theta_3\pm\theta_4 \;\;\;\ 0r\;\;\;\ \theta_1+\theta_3=\theta_2+\theta_4.
\ee
Hence the amplitude vanishes to all orders of $t$ when (8) is satisfied, 
i.e. in the case that branes make 4 angles related by (8), a certain fraction of 
SUSY must remain. We will argue in section 4 that, this configuration in general 
(for an arbitrary choice of angles $\theta_i$ upto the condition (8) ) preserves 
${1 \f 16}$ of SUSY, which has been sought for [16]. 

The small $t$ limit of the integrand of (4), which shows the contribution of the 
massless closed strings exchanged between branes, reads to be:
\be
A(\theta_i) =V_0[\sum_{i=1}^4 cos 2\theta_i - 4\prod_{i=1}^4 cos \theta_i +
 4\prod_{i=1}^4 sin \theta_i],
\ee
where $V_0$ is proportional to $T_p^2=(4\pi^2\alpha')^{3-p}$. The first term
in the bracket is graviton and dilaton contribution and rest is RR contributions.
Now let us consider some important special cases which reproduces the 
interaction configuration with less number of angles associated with the 
families with more SUSY.

{\it special cases}

a) $\theta_1=\theta_2=\theta_3=\theta_4=0 \;\;\;\;\;\;\; ,A_0(0)=0$,
which recovers results of [1].

b) $\theta_2=\theta_3=\theta_4=0\;\;\;\ ,  
A_1(\theta_1)=2V_0(1-cos\theta_1)^2$, in which the RR contribution is
proportional to $-cos\theta_1$, and graviton and dilaton proportional to  
$(1+cos^2\theta_1)$. This is what discussed in detail in [17].

c)  $\theta_3=\theta_4=0 \;\;\;\;\;\;\;  ,A_2(\theta_1,\theta_2)
=2V_0(cos\theta_2-cos\theta_1)^2$.

d) $\theta_4=0 \;\;\;\;\;\; ,A_3(\theta_1,\theta_2,\theta_3)=
2V_0[cos(\theta_1+\theta_2)-cos\theta_3][cos(\theta_1-\theta_2)-cos\theta_3] $.  

e)  $\theta_1=\theta_2=\theta_3=\theta_4=\pi/2 \;\;\;\;\; ,A_4(\pi/2)=0$.

f)  $\theta_2=\theta_3=\theta_4=\pi/2 \;\;\;\;\; ,A_5(\theta_1)=
2V_0(1-sin\theta_1)^2$.

g)  $\theta_3=\theta_4=\pi/2 \;\;\;\;\;\;\; ,A_6(\theta_1,\theta_2)
=2V_0(sin\theta_2-sin\theta_1)^2$.

h) $\theta_4=\pi/2 \;\;\;\;\;\;\;\ ,A_7(\theta_1,\theta_2,\theta_3)=
2V_0[cos(\theta_1+\theta_2)-sin\theta_3][cos(\theta_1-\theta_2)-sin\theta_3]$.  

As we see $A_1,\ A_5$ vanish only for $\theta_1=0\;\;or \;\ \pi/2$  respectively, 
so we have not a stable brane configuration 
related by only one independent rotation of $U(4)$.
$A_2,\ A_6$ vanish for $\theta_1=\theta_2$ and $\theta_1=-\theta_2$ (for $A_2$)
and $\theta_1=\pi-\theta_2$ (for $A_6$). These cases make a one parameter family
of stable branes.
$A_3,\ A_7$ vanish for $\theta_1\pm\theta_2=\pm \theta_3$ and 
$\theta_1\pm\theta_2=\pi/2 \pm \theta_3$ respectively. They describe a two 
parameter family of stable branes.
All the above families are special cases of three parameter families given 
by (8).
\vskip 0.5cm
{\bf 3.   $(p,p')$ branes at angles}

Untill now we have only considered two $D_p$-branes case. In this section, we
show that $(p,p')$ branes making angles can be studied as special case of the 
general 4 angles case.

Let us denote $p-p'=\Delta\;\;\; (p>p')$; parallel $(p,p')$ branes can be 
introduced by the following boundary conditions:

\be
\sigma=0 \left\{  \begin{array}{cc}
X^{\mu}=0 & \mu=p'+1,..,9 \\ 
\partial_{\sigma}X^{\mu}=0  &  \mu=0,..,p'
\end{array}\right.
\ee
\be
\sigma=\pi \left\{  \begin{array}{cc}
X^{\mu}=Y^{\mu} & \mu=p+1,..,9 \\ 
\partial_{\sigma}X^{\mu}=0  &  \mu=0,..,p
\end{array}\right.
\ee
which for $\Delta=2$ is exactly the boundary conditions of (1),(2) for 
$\theta_1=\pi/2 ,\ \theta_2=\theta_3=\theta_4=0$, generally $\Delta=2n$,
is equivalent to
\be
\left\{ \bea{cc}
\theta_i=\pi/2  \;\;\;\; i \leq n\\
\theta_i=0 \;\;\;\; otherwise. 
\eea\right.
\ee

So all the results of $(p,p')$ branes at angles can be obtained from the 4 angle
calculations. More precisely their amplitude are different only in the tension 
coefficient which is $T_pT_{p'}$ for $(p,p')$ and $T_p^2$ for $(p,p)$ case.
Hence the most general case of rotated $\Delta=2$ amplitude is proportional to
$A_7$, for $\Delta=4,6$ proportional to $A_6,A_5$ respectively, in the spcial
cses discussed above. 
For $\Delta=8$ case the amplitude vanishes which is described by $A_4$ case.
\vskip 0.5cm
{\bf 4.   SUSY arguments}

In this section first we briefly review the SUSY constraints:
\newline
As we know any p-brane  which its world volume spans $012..p$ preserves $1\f 2$ 
of 32 super charges of type II or M-theory. For D-branes of type II they are 
associated with constraints
\footnote{ metric is (-- + ....++) and 
$\Gamma_0=\left(\bea{cc} 
0\;\ -1\\
1\;\;\ 0
\eea\right)$ and 
$\Gamma_i=\left(\bea{cc} 
0\;\ \gamma_i\\
\gamma_i\;\;\ 0.
\eea\right)$, where $\gamma_i$ are real symmetric matrices.}
\be
\Gamma_{01..p}\eps_L=\eps_R,
\ee
where $\eps_L,\eps_R$ are parameters of left and right moving spinors of
type II theories.

Any rotated $D_p$-brane must satisfy the condition
\be
(R\Gamma_{01..p} R^{-1}) \eps_L=\eps_R.
\ee
So each $\eps_L$  satisfied both (13) , (14) gives the fraction of SUSY which
is preserved. In our case, which $U(p)$ rotations of $SO(2p)$ is considered
\be
R\Gamma_{01..p} R^{-1}=R^2 \Gamma_{01..p}.
\ee
Then (14) can be replaced by 
\be
R^2 \eps=\eps.
\ee
For 4 angle rotations described in section 2 (($U(1))^4$ members of $U(4)$)
the rotation $R^2$ is written as
\be
R^2=exp(\sum_{i=1}^4 \theta_i  \Gamma_{i,4+i})=\prod_{i=1}^4 (cos\theta_i+
\Gamma_{i,i+4}sin\theta_i).
\ee
Carrying out the calculations, we find that $(R^2-1)$ has zero eigen-values
if and only if
\be
\Gamma_{i,i+4}\Gamma_{j,j+4}\eps=\eps \;\;\;\;\; ,i\neq j,
\ee
which yields  {\it the conditions of (8) on $\theta_i$ angles}, required for
the vanishing of the interaction.
The number of conserved super charges then can be obtained from (18) equations, 
in the most general from these equations can be explicitly written as
\be
\left\{  \bea{cc}
A\eps=\eps\;\;\;\;\ , A=\Gamma_{1526} \\
B\eps=\eps\;\;\;\;\ , B=\Gamma_{1537} \\
C\eps=\eps\;\;\;\;\ , C=\Gamma_{1548}.
\eea\right.
\ee
Since the matrices $A,B,C$ commute with each other, are traceless and their
square is equal to  
${\bf 1}$; and also as $AB, AC$ are traceless, the above equations will be 
satisfied simultaneously only for one $\eps$ out of 16.

Hence in general the 4 angle contribution provided that (8) holds, preserves
${1 \f 16}$ of SUSY. There are some special cases such as $\theta_1=\theta_2,
\theta_3=\theta_4$ which allows more SUSY, i.e. ${1 \f 8}$.

Although we have done our calculations for D-branes in strings theory, the 
results on angles can be extended to the other kinds of branes ($NS_5$ or 
$M_5$- branes) through SUSY arguments, i.e. two $NS_5$ or $M_5$-branes at 
angles satisfying (8) preserves ${1 \f 16}$ of SUSY.

Besides SUSY algebraic arguments, which are true for NS or M five branes, 
dualities can also relate configurations containing $NS_5$ or $M_5$-branes to  
those of D-branes and hence our D-branes results are generalized to these cases.
Corresponding dualities are T-duality of IIA, IIB theories. Under this duality
a stable system of D-branes goes to another stable configuration which has the 
same number of super charges. 
If we T-dualize a brane in an arbitrary direction $D_p$-brane goes to bound state 
of F-string with a $D_{p+1}$-brane [3], which is a non-marginal bound state
and preserves the same SUSY fraction as $D_p$-brane does, i.e. ${1 \f 2}$. 
So we can find configurations of two branes in which each brane is a bound state
of F-string and D-brane.
The same argument holds for Hodge duals( electric, magnetic duals) of 
the above configurations where every bound state of F-string and $D_p$-brane
goes to $NS_5$ and $D_{6-p}$-brane bound state. 

In type IIB theory there are also $(m,n)$string or five brane bound states
[2] which preserves ${1 \f 2}$ of SUSY (Such branes are non-marginal bound 
states).

The M-theory branes ($M_2$,$M_5$-branes) can be understood as U-duals of IIA branes.
This duality like, T-duality of IIA and IIB preserves number of conserved super 
charges, so one expects to find non-marginal bound states of $M_2$, $M_5$-branes
(duals of $D_4$-brane, F-string bound state) with ${1 \f 2}$  of SUSY [15].
Other $M_5$-brane configurations are duals of $D_4$-brane configurations we 
have considered here.
\vskip 0.5cm
{\bf 5.   Summery}

In this section, we summerize the results obtained  partly by others (mainly in
[5,16]) and partly in this article ordered in the fraction of SUSY. 
Unless it is mentioned explicitly, the results are true for D-branes, $NS_5$-branes
and $M_5$-branes.
\vskip 1cm
{\it i)${1 \f 2}$ Supersymmetry}
\vskip 0.5cm

Parallel branes or non-marginal bound states of ($D_p$-branes, F-strings) and
($NS_5$-branes, D-branes) and ($M_5$-branes and $M_2$-branes). Also KK-monopoles 
of type II or  M-theory on $S^1$ [18].
\vskip 1cm
{\it ii)${1 \f 4}$ Supersymmetry}
\vskip 0.5cm

In angles arguments:

$\theta_1=\theta_2=\theta_3=\theta_4=\pi/2. \hspace{3cm}
\;\;\;\; \mid \;\;\; \theta_1=\theta_2\;\;\ ,\;\;\;\theta_3=\theta_4=0.$ 

$\Delta=2,\;\;\;\ \theta_1=\theta_2=0\;\;\;,\;\;\; \theta_3=\pi/2. \hspace{1.5cm}
\mid \;\;\; \Delta=2,\;\;\;\ \theta_1=\theta_2=\theta_3=\pi/2.$

$\Delta=4,\;\;\;\ \theta_1=\theta_2=0. \hspace{4cm} \mid \;\;\
\Delta=4,\;\;\;\ \theta_1=\theta_2=\pi/2.$ 

$\Delta=6 \;\;\;\;\ \theta_1=\pi/2. \hspace{4.5cm} \mid  \;\;\;   
\Delta=8.$

$NS_5,\;\ D_4$: $\theta_1=\theta_2=\theta_3=0\;\; , \theta_4=\pi/2$ in type IIA [19].

$NS_5,\;\ D_3$: $\theta_1=\theta_2=0\;\; , \theta_3=\pi/2$ in type IIB [20].

$NS_5,\;\ D_2$: $\theta_1=0 \;\;\ , \theta_2=\pi/2$ in type IIA  and

$M_5,\;\ M_2$: $\theta_1=0 \;\;\ , \theta_2=\pi/2$ in M-theory level [21].

$NS_5,\;\ D_1$: $\theta_1=\pi/2$ in type IIB [22].

\vskip 1cm
{\it iii)${3 \f 16}$ Supersymmetry}
\vskip 0.5cm

$\theta_1=\theta_2=\theta_3=\theta_4 \neq \pi/2. \hspace{3cm}$

\newpage
{\it iv)${1 \f 8}$ Supersymmetry}
\vskip 0.5cm

$ \theta_1\pm \theta_2=\pm \theta_3\;\;\;\ ,\theta_4=0. \hspace{3.45cm} \mid
\;\;\;\; \theta_1=\theta_2\neq 0\;\;\ , \theta_3=\theta_4 \neq 0.$ 

$\Delta=2,\; \theta_1=\theta_2\neq 0\; ,\; \theta_3=\pi/2. 
\hspace{2.3cm}\mid \;\;\;  \Delta=2,\; \theta_1=\pi-\theta_2\;\; ,
\theta_3=\pi/2.$

$\Delta=4,\;\;\;\ \theta_1=-\theta_2. \hspace{4.35cm} \mid
\;\;\;\;\Delta=4,\;\;\;\ \theta_1=\pi-\theta_2.$ 

\vskip 1cm
{\it v)${1 \f 16}$ Supersymmetry}
\vskip 0.5cm

$\theta_1+\theta_2+\theta_3+\theta_4=0. \hspace{3.6cm}
\;\;\;\; \mid \;\;\; \theta_1+\theta_2=\theta_3+\theta_4.$ 

$\Delta=2,\;\;\;\ \theta_1+\theta_2+\theta_3+\pi/2=0. \hspace{2cm}
\mid \;\;\; \Delta=2,\;\;\;\ \theta_1+\pi/2=\theta_2+\theta_3.$

\vskip 1cm

{\bf Appendix}:
\vskip 0.5cm
\cl{{\bf New $\Theta$-function identity}}  

Here we present a proof for the following identity between Jacobi 
$\Theta$-functions:

$$
\prod_{j=1}^{4}\Theta_3(\nu_j  \mid \tau)
-\prod_{j=1}^{4}\Theta_4(\nu_j \mid \tau)
=
\prod_{j=1}^{4}\Theta_2(\nu_j  \mid \tau)
-\prod_{j=1}^{4}\Theta_1(\nu_j \mid \tau)
$$ 
provided that $ \sum_i \nu_i=0$ or sum of any two of them is equal two others.

To prove we denote that: 
1) Each $\Theta$-function is doubly periodic with respect to $\nu_i$. 

2) It has certain zeros and poles.

As it is mentioned in [23,24] in order to prove the identity it is enough to 
check both sides first, to be doubly periodic (with periods to 1, $\tau$)
with respect to $\nu_i$, second, to have same zeros and poles, then the ratio 
of both sides is a constant and the value of this constant is determined by using 
special values of $\nu_i$ (e.g. $\nu_i=0$). So we check these conditions:

1){\it periodicity:}

Let
$$
\Theta_a(\nu_1 \mid \tau)\Theta_a(\nu_2 \mid \tau)
\Theta_a(\nu_3 \mid \tau)\Theta_a(\nu_1+\nu_2+\nu_3 \mid \tau)\equiv F_a
$$
Under $\nu \rightarrow \nu+1$, $F_a$ is not changed for $a=1,2,3,4$.

Under $\nu \rightarrow \nu+\tau$, $F_a$ goes to $A^4(\nu)F_a$, where
$A(\nu)=e^{-i\pi(2\nu+\tau)}$. 

Hence the identity is doubly periodic with periods $(1,\tau)$.

2){\it Zeros and Poles:}

The poles can be checked in small $i\tau$ limit $(q=e^{-i\pi\tau}\approx 1)$.
This limit could be studied by modular transformation on large $i\tau$ limit
$(q\approx 0)$. The $\nu$ dependence of this limit is proportional to the four 
angle amplitude (9) which is true when the corresponding relation between 
$\nu_i$ holds.

The zeros can be checked as following. Let us consider the identity as: 
$$
F_2+F_4-F_1=F_3.
$$
We check that for every root of the right hand side, left hand side vanishes. 
If we also check that for any root of $F_2$ the identity holds, then we can 
be sure about a one to one correspondence between zeros. As we know $\Theta_3$
vanishes at $\nu=m+{1\f 2}+(n+{1\f 2})\tau\;\;\, m,n\in Z$ [23]. Replacing 
this in left hand side and using half period behaviour of $\Theta$-functions 
we find that left hand side vanishes provided that the following identity holds:
$$
\Theta_2(0 \mid \tau)\Theta_2(\nu_1 \mid \tau)
\Theta_2(\nu_2 \mid \tau)\Theta_2(\nu_1+\nu_2 \mid \tau)=
$$
$$
\Theta_3(0 \mid \tau)\Theta_3(\nu_1 \mid \tau)
\Theta_3(\nu_2 \mid \tau)\Theta_3(\nu_1+\nu_2 \mid \tau)-
\Theta_4(0 \mid \tau)\Theta_4(\nu_1 \mid \tau)
\Theta_4(\nu_2 \mid \tau)\Theta_4(\nu_1+\nu_2 \mid \tau).
$$

which is given in [24] ( The above identity can also be proved by the 
same method we presented here: 1) checking the periodicity with periods 
$(1, \tau)$. 2) Proving the one to one correspondence between zeros and poles.).
So the ratio of the both sides is constant, if we put all $\nu_i$ to be zero
we find that this constant is one.   QED.  
\vskip 0.5cm

{\bf Acknowledgement:}

Author would like to thank H. Arfaei for fruitful discussions.

\end{document}